\documentclass[sigconf,10pt]{acmart}

\setcopyright{none}
\renewcommand\footnotetextcopyrightpermission[1]{}
\usepackage{xcolor}

\usepackage{enumitem}

\newcommand{\firepdfauthors}{Shripad Nadgowda, Gaurav Mitra, Sing Sing Ma, Vipul Chaskar, Pradipta Ghosh, Canturk Isci, Onkar Deshpande, Edgardo Aviles-Lopez, Ryan Feng, Ján Ondrušek}

\begin{document}

\title{FIRE: Failure Isolation and Reliability for Production-Connected Experiments at Scale}

\settopmatter{authorsperrow=4}

\author{Shripad Nadgowda}
\affiliation{\institution{Meta, Inc.}\country{USA}}
\email{shri.nadgowda@gmail.com}

\author{Gaurav Mitra}
\affiliation{\institution{Meta, Inc.}\country{USA}}
\email{gmitra@meta.com}

\author{Sing Sing Ma}
\affiliation{\institution{Meta, Inc.}\country{USA}}
\email{singsing@meta.com}

\author{Vipul Chaskar}
\affiliation{\institution{Meta, Inc.}\country{USA}}
\email{vchaskar@meta.com}

\author{Pradipta Ghosh}
\affiliation{\institution{Meta, Inc.}\country{USA}}
\email{iampradipta@meta.com}

\author{Canturk Isci}
\affiliation{\institution{Meta, Inc.}\country{USA}}
\email{cisci@meta.com}

\author{Onkar Deshpande}
\affiliation{\institution{Meta, Inc.}\country{USA}}
\email{odeshpande@meta.com}

\author{Edgardo Aviles-Lopez}
\affiliation{\institution{Meta, Inc.}\country{USA}}
\email{edgardo@meta.com}

\author{Ryan Feng}
\affiliation{\institution{Meta, Inc.}\country{USA}}
\email{zhengliangfeng@meta.com}

\author{J\'an Ondru\v{s}ek}
\affiliation{\institution{Meta, Inc.}\country{USA}}
\email{ondrusek@meta.com}

\renewcommand{\shortauthors}{Nadgowda et al.}

\begin{abstract}
  Pre-production changes are usually tested in isolated environments to prevent
  their side effects from reaching production. Some record-and-replay platforms
  allow a production-connected mode: candidate changes receive recorded production
  requests while selected downstream calls reach production dependencies. This
  improves fidelity before deployment, but it also introduces
  reliability risks. At hyperscale, experimental traffic can fan out across shared
  services, trigger latent defects, or send malformed input to systems serving
  user requests.

  In this paper, we show how production-connected experimentation can operate reliably in a large-scale recommendation serving system.
  To contain downstream effects without sacrificing realistic dependency behavior, the platform uses three complementary guardrails.
  Isolated tiers redirect replay traffic away from shared
  dependencies where replication is practical. Dependency monitors observe
  downstream health and experiment-attributed errors and terminate experiments
  after sustained violations. Traffic pinning limits experimental calls to a
  designated subset of downstream workers when full replication would require
  scarce capacity.

  Isolated ranking tiers process billions of replayed requests per day using under
  1\% of the corresponding serving capacity. Over 90 days, these tiers experienced
  more than 200{,}000 abnormal worker exits without directly affecting
  production-serving workers. During a separate three-month window, dependency
  monitors cancelled approximately 5\% of experiments for the measured ranking
  service and accounted for 87\% of its cancellations. Dependency-aware triggering
  also reduced code-review experiment executions by 25\%. These results show that
  high-fidelity, production-connected experimentation can operate at scale when
  downstream effects are isolated where possible, bounded where necessary, and
  continuously monitored and attributed.
\end{abstract}

\maketitle
\pagestyle{plain}
\hypersetup{pdfauthor={\firepdfauthors}}
\pdfstringdef\firepdfauthorstring{\firepdfauthors}
\pdfinfo{/Author (\firepdfauthorstring)}

\section{Introduction}
  \label{sec:intro}

  Testing candidate changes before deployment is an established part of software
  delivery. These tests are typically run in environments isolated from production
  so that crashes, malformed requests, and excess resource use cannot affect production
  systems. Isolation, however, can weaken the resulting signal. Mocked dependencies
  may omit important cross-service behavior, while complete replicas require current
  data, configuration, and specialized hardware. Canary testing provides realistic
  dependency behavior, but only after a change has entered the release path.

  Record-and-replay platforms improve pre-deployment fidelity by exercising
  candidate code with requests captured from production
  \cite{liu2022recordreplay,diffy}. Some platforms also permit selected calls from
  the experimental deployment to reach production dependencies
  \cite{chow2024servicelab}. We call this model \emph{production-connected
  experimentation}. It combines an isolated service under test (SUT) with realistic
  downstream behavior before the candidate change lands.

  Operating this model for a hyperscale recommendation serving system creates a
  separate reliability problem. A recommendation request can cross query,
  aggregation, ranking, feature, inference-gateway, and prediction services and fan
  out to many downstream workers. Experimental traffic can therefore consume shared
  capacity, exercise latent dependency defects, or carry malformed input beyond the SUT.
  Failures may appear several service boundaries away from the experiment that
  caused them, making containment and attribution difficult.

  Observed failures illustrate this problem. In one case, replayed requests
  repeatedly exercised a latent timeout defect in a shared ranking service and
  caused workers to restart. In another, a candidate ranking change sent malformed
  tensor metadata to a shared prediction service and left accelerator-backed
  workers unable to serve requests. Other effects included delayed resource cleanup
  and contamination of downstream telemetry. These cases show that protecting the
  SUT is insufficient when its outgoing calls remain connected to shared services.

  The platform combines three guardrails. \emph{Isolated tiers} redirect replay
  traffic away from shared dependencies where a representative replica can be
  maintained. \emph{Dependency monitors} observe downstream health and
  experiment-attributed errors, then terminate experiments after sustained
  violations. \emph{Traffic pinning} limits experimental calls to a designated
  subset of downstream workers when full replication would require scarce capacity.
  The mechanisms provide different guarantees and are selected according to each
  dependency's state, capacity, and failure modes.

  Scale also makes unnecessary experiments a reliability concern because every run
  consumes platform resources and may add load to shared dependencies. The platform
  therefore uses build-graph reachability for source changes and observed runtime
  consumption for configuration changes. This dependency-aware triggering reduced
  code-review experiment executions by 25\%.

  This paper makes three contributions:

  \begin{itemize}
    \item We characterize the reliability risks created when pre-deployment
    experiments retain access to shared production dependencies.

    \item We present a guardrail design that combines selective isolation,
    continuous downstream monitoring and attribution, and bounded worker exposure.

    \item We evaluate these mechanisms at scale. Isolated ranking tiers process
    billions of replayed requests per day using under 1\% of the corresponding
    serving capacity and experienced more than 200{,}000 abnormal worker exits over
    90 days without directly affecting workers serving production traffic. Dependency
    monitors accounted for 87\% of experiment cancellations in a separate
    three-month window.
  \end{itemize}

  These results show how production-connected experimentation can provide realistic
  pre-deployment signals while keeping downstream effects observable, attributable,
  and contained.

\section{System Context}
\label{sec:context}

\subsection{Recommendation serving stack}
\label{sec:context:stack}

\begin{figure}[t]
\centering
\includegraphics[width=0.9\columnwidth]{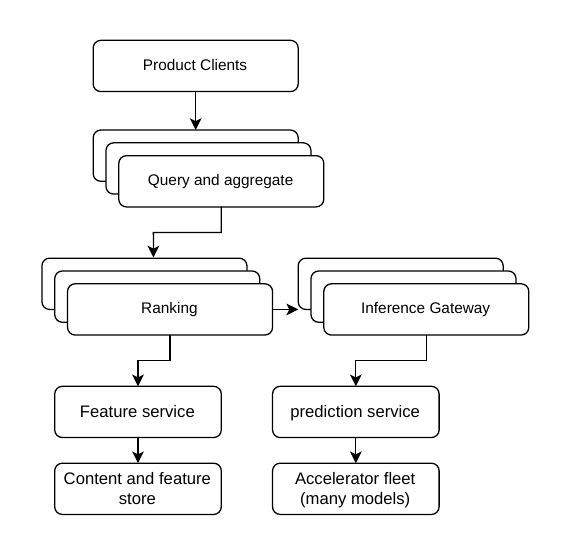}
\caption{Representative serving graph scoped to this paper. A request enters
query and aggregation, passes to ranking, and fans out to a feature service and an
inference gateway. The gateway routes model requests to a prediction service backed
by an accelerator fleet. Stacked outlines indicate independent tiers.}
\Description{A product request enters a query and aggregation service and then a
ranking service. Ranking calls a feature service backed by a content and feature
store, and an inference gateway that routes to a prediction service backed by an
accelerator fleet. Stacked boxes indicate independent service tiers.}
\label{fig:stack}
\end{figure}

In the recommendation serving stack, each request traverses a graph of cooperating
services rather than a single binary. Figure~\ref{fig:stack} shows the
representative part of that graph that is in scope for this paper.

A \emph{query and aggregation service} receives the product request, resolves
  viewer and surface context, assembles the candidate set to be scored, and
  combines ranked results into the response returned to the product. A
  \emph{ranking service} scores and orders those candidates, retrieving model
  features from feature services and sending model-evaluation requests through an
  \emph{inference gateway}. The gateway routes those requests to a \emph{prediction
  service}, which executes machine-learned models on accelerator-backed workers.

The request graph widens at ranking. Scoring one candidate set can require many
  calls to feature services and the inference gateway. The prediction layer hosts
  many models because different request classes and stages of the ranking funnel
  use different model sets. Query and aggregation, ranking, and the inference
  gateway are independently deployed services and can each be selected as the SUT.

Two properties matter for production-connected experimentation. A failure can
surface several service boundaries away from the component that caused it, and
one ranking request can issue many downstream calls. Replaying a small request
stream can therefore create meaningful demand on a shared dependency.

We use \emph{product surface} to mean a user-facing feature that requests
recommendations, such as a feed or short-form video view. Recommendation services commonly
run as multiple independent \emph{tiers}. A tier is an instance group for a
particular product surface or request path.
  Tiers may share code while differing in request mix, candidate-set size,
  dependency fan-out, model set, and latency budget. A change that is neutral for one tier can therefore regress another, so validation is organized around the tier being exercised rather than only around
  the service binary.

The dependency classes resist isolation for different reasons. These differences
determine which containment mechanism applies in
  Section~\ref{sec:guardrails}. Ranking and inference-gateway tiers can often be
  replicated as isolated tiers because their serving state and configuration can
  be kept close enough to the corresponding shared tier for replay comparisons to
  remain meaningful. Feature services are harder to replicate because they depend
  on large, continuously updated stores; stale or divergent state can cause the
  same replayed request to produce different results for reasons unrelated to the
  candidate change. Prediction services are constrained by accelerator-backed
  capacity and by the number of distinct model instances. As a result, no single
  containment strategy applies uniformly across the graph.

\paragraph{Experiment model.}
The platform selects one service tier as the SUT, runs control and treatment
versions in comparable environments, and replays the same sampled requests
against both. A \emph{workload} binds that tier to its request corpus, resource
reservation, and counters. The SUT receives no user requests, but downstream calls
that are not redirected can still reach shared dependencies.

\section{Real-World Failure Modes}
\label{sec:incidents}

Production-connected experimentation creates failure modes that do not appear in
  fully isolated testing. Replayed requests can exercise shared downstream
  dependencies, so the visible failure may occur outside the SUT that originated
  the traffic. The two scenarios below show complementary causes: replay can expose
  a latent defect in a dependency, or a candidate change can send invalid input to
  that dependency. Table~\ref{tab:risks} summarizes the broader risks and the
  platform response to each.

\begin{table*}[t]
  \centering
  \small
  \setlength{\tabcolsep}{3pt}
  \begin{tabular}{@{}p{0.28\textwidth}p{0.32\textwidth}p{0.28\textwidth}@{}}
    \toprule
    Risk & Consequence & Platform response \\
    \midrule
    Replay reaches a replicable dependency
      & Shared serving capacity can be affected
      & Isolated tier \\
    Failure appears outside the SUT
      & Its cause is difficult to identify
      & Dependency monitor and request identity \\
    Fault reaches many workers
      & Small traffic volume can cause broad impact
      & Traffic pinning \\
    Old replay keeps dependency state active
      & Cleanup or model reaping is delayed
      & Bounded replay lookback \\
    Unrelated workloads are executed
      & Resource use and downstream exposure increase
      & Dependency-aware triggering \\
    \bottomrule
  \end{tabular}
  \caption{Risks in production-connected experimentation and the platform
    response to each.}
  \label{tab:risks}
\end{table*}

\subsection{Exercising a latent dependency defect}
\label{sec:incidents:latent}

One scenario began with a change to the ranking service's request-timeout
  handling. On timeout, the service canceled the request but allowed an
  asynchronous feature-extraction task within the request handler to continue
  running. That task still held references to request-scoped objects, so after the
  request was torn down it could de-reference freed state and crash the worker. The
  change passed pre-submit validation because the timeout path depended on a new
  caller pattern that was rarely exercised by ordinary request traffic at the time
  of submission.

A record-and-replay experiment later drove recorded requests through that path
  at experiment rate. The replay did not introduce the defect, but it concentrated
  enough matching requests to exercise the latent path repeatedly. Ranking workers receiving the experiment's replayed traffic began crashing and restarting. Automated health signals
  detected the failures quickly, but the first diagnostic step was attribution:
  the restart signature identified the failing service, not the upstream
  experiment whose replayed requests were triggering the failure. Containment therefore required separating experiment-originated traffic from
  ordinary requests and then stopping the relevant experiments. Stopping the
  experiments removed the replay pressure, while the timeout path was fixed
  separately so the asynchronous task could not outlive the request state it
  accessed.

This scenario illustrates why reproducing downstream behavior is different from
  containing downstream effects. The same mechanism that creates risk also creates
  signal: replayed requests exercise real downstream behavior, but any effect of
  those requests can also reach the shared dependency. The replayed requests were
  sampled from ordinary serving traffic, but the defect was timing-dependent
  rather than content-dependent. It appeared only when a request timed out after
  asynchronous feature-extraction work had started but before that work finished.
  Replay can change that timing by repeating sampled requests and driving the SUT
  at a controlled utilization, which changes queuing, cache behavior, and
  downstream response timing. As a result, requests that completed normally when
  originally served can exercise a cancellation interleaving during replay. The
  reliability concern arose because those replayed requests reached a shared
  downstream dependency rather than an isolated equivalent; where such isolation is
  impractical, the platform needs monitoring and attribution at the downstream
  dependency.

\subsection{Injecting invalid input}
\label{sec:incidents:invalid}

A contrasting scenario starts with the candidate change rather than with a latent
  defect in the dependency. A ranking change altered how features were converted
  into tensors before calling a shared prediction service. For one model
  configuration, the change treated a per-candidate feature as though it were
  request-level state, collapsing the batch dimension. The resulting tensor had
  shape $[1, \mathit{features}]$ where the model contract required
  $[\mathit{batch}, \mathit{features}]$.

The malformed tensor crossed the SUT boundary and failed inside the prediction
  service. The failure was not a local ranking error: the downstream service
  received input that violated a model contract, triggered an accelerator
  assertion, and left affected workers unable to continue serving requests until
  they were restarted. The visible symptoms therefore appeared in the shared
  dependency rather than in the deployment that produced the malformed input.

Diagnosis required correlating evidence from multiple services: the ranking
  experiment that constructed the tensor, the inference gateway that routed the
  request, and the model whose input contract was violated. The first response was
  to stop the experiment and recover the affected workers.

The same failure mode later appeared through other request paths to the shared
  dependency. The difficult part was identifying all upstream services and request
  paths that could emit the malformed input. This recurrence showed that coverage
  is part of the design problem. A mechanism protects only the request paths and
  dependencies to which it is applied.

\subsection{What the pattern requires}
\label{sec:incidents:lessons}

  The scenarios above share three properties, each of which maps to a guardrail in
  Section~\ref{sec:guardrails}.

  \emph{Detection must be paired with attribution.} Health signals can show that a
  downstream dependency is failing, but they do not automatically identify which
  experiment caused the failure. When routing, error, or restart records do not
  carry experiment identity, separating experiment-originated requests from other
  requests becomes a diagnostic step rather than a query. A guardrail that
  preserves caller identity is therefore the precondition for stopping one faulty
  experiment without interrupting unrelated validation.

  \emph{Containment must account for fan-out.} Experimental traffic may be small
  relative to overall dependency volume, but crash-class failures are not
  necessarily proportional to traffic share. If an experiment fans out across many
  workers, one malformed request per worker can make the affected set large even
  when the experiment's request volume is modest. For these failures, the relevant
  quantity is the number of workers reachable by replayed traffic, not only the
  fraction of requests the experiment contributes.

  \emph{SUT-local validation is insufficient.} Pre-submit checks can correctly
  evaluate the SUT while missing effects that appear only after the SUT calls a
  shared downstream dependency. The failure may depend on request timing,
  downstream batching, model contracts, or dependency state outside the SUT's
  local counters. The platform therefore needs guardrails that observe downstream
  dependencies directly and connect their signals back to the originating
  experiment.

  Together, these requirements define where each guardrail applies. Isolation
  removes a shared edge where a representative tier is practical. Monitoring
  covers dependencies that remain shared. Pinning limits fan-out for failures whose
  impact depends on the number of workers reached. Section~\ref{sec:guardrails}
  describes these choices and their remaining limits.

\section{Downstream Guardrails}
\label{sec:guardrails}

\begin{figure*}[t]
  \centering
  \includegraphics[width=0.72\textwidth]{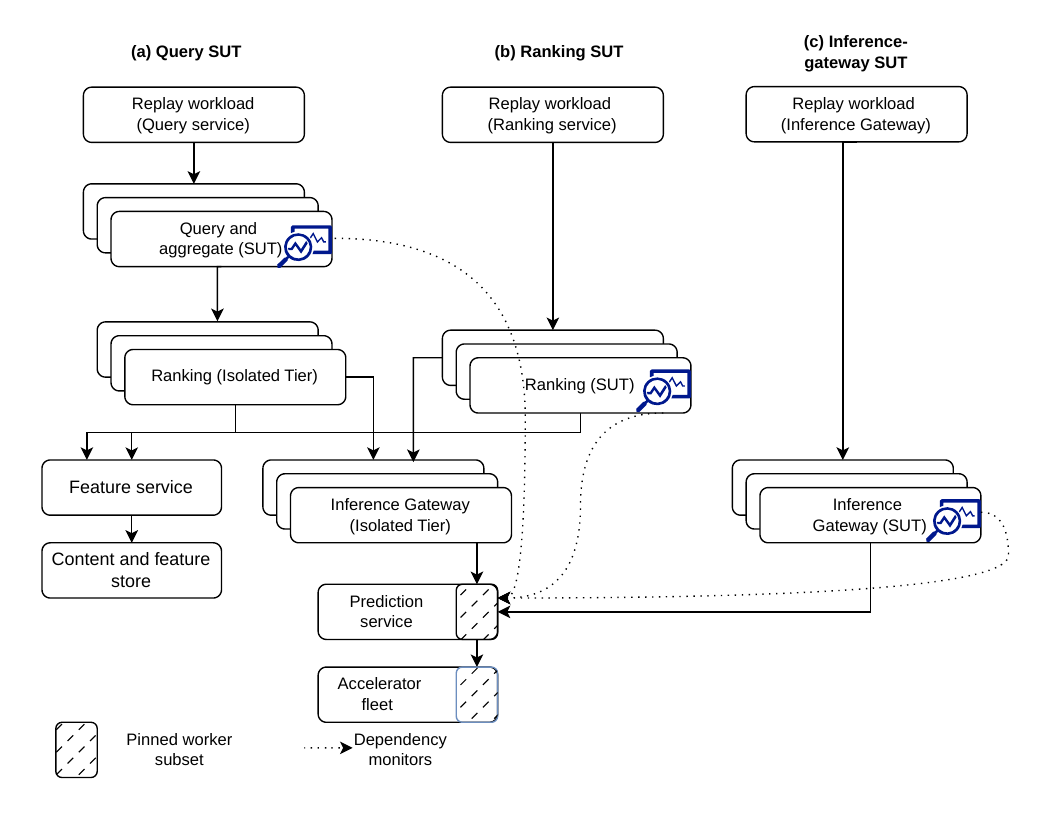}
  \caption{Guardrail placement as the SUT moves downstream. (a) A query
  experiment routes calls through isolated ranking and inference-gateway tiers.
  (b) A ranking experiment uses an isolated inference-gateway tier. (c) An
  inference-gateway experiment reaches the shared prediction service directly.
  Dependency monitors observe the prediction service in all three examples, and
  traffic pinning limits experimental calls to the hatched worker subset. Solid
  arrows show requests; dotted arrows show monitoring.}
  \Description{Three experiment paths share a prediction service. A query SUT
  calls isolated ranking and inference-gateway tiers. A ranking SUT calls an
  isolated inference-gateway tier. An inference-gateway SUT calls prediction
  directly. Dotted lines show dependency monitoring, and hatched regions mark the
  pinned prediction workers and accelerators.}
  \label{fig:guardrails}
\end{figure*}

This section presents three guardrails for containing downstream effects of
  production-connected experiments: isolated tiers, dependency monitors, and
  traffic pinning. Figure~\ref{fig:guardrails} places these mechanisms on the
  representative serving graph and shows where each acts on the downstream path.

\subsection{Selective isolation}
\label{sec:guardrails:isolation}

The strongest guardrail is to route replayed downstream calls away from the
  shared dependency entirely. We use \emph{isolated tiers} for dependencies
  that can be replicated with enough fidelity for replay comparisons to remain
  meaningful. When such a tier exists, experiment traffic reaches the isolated
  tier rather than the shared one, so failures in that dependency
  class are contained within the experimental path.

Isolated tiers are provisioned selectively because replicating the full
  serving graph is not practical. Three criteria determine where they are useful.
  First, the dependency's serving state and configuration must be reproducible
  enough that the same replayed request produces comparable behavior; a stale or
  divergent replica would make the control-treatment comparison inconclusive.
  Second, the tier must be maintainable as the corresponding shared tier changes,
  including software, configuration, scaling, and health
  checks. Third, the required capacity must be practical to reserve. This is why
  isolated tiers are more suitable for ranking than for accelerator-backed
  prediction models or stateful feature stores.

Selected ranking and inference-gateway deployments satisfy these criteria and can
  be maintained as isolated tiers. Replayed calls that would otherwise reach their
  shared counterparts are routed to the corresponding isolated tier. For the
  ranking service measured in Section~\ref{sec:eval}, these isolated tiers use, in
  aggregate, less than one percent of the worker capacity of the corresponding
  shared ranking tiers. Isolation removes that downstream exposure rather than
  merely bounding it. The tradeoff is resource demand: it is preferred where a
  representative tier can be kept current, but it is not a general solution for
  every dependency in the serving graph.

\subsection{Continuous detection: dependency monitors}
\label{sec:guardrails:monitoring}

When a downstream dependency remains shared, the platform monitors it during the
  experiment. A \emph{dependency monitor} runs alongside the experiment, polls a
  configured set of detectors, and terminates the experiment when a detector
  reports a sustained violation.

  \begin{figure}[t]
  \small
  \begin{tabbing}
  \hspace*{1.2em}\=\hspace*{1.2em}\=\hspace*{1.2em}\=\kill
  \textbf{for} each detector $d$ configured for the workload: \\
  \> $d.\textsf{prime}()$; \quad $c_d \gets 0$ \\[2pt]
  \textbf{while} the experiment is running: \\
  \> sleep until the next detector is due \\
  \> \textbf{for} each due detector $d$, with $t$ the elapsed time: \\
  \> \> \textbf{if} $t < \mathit{warmup}_d$: \textbf{continue} \\
  \> \> \textbf{if} $t > \mathit{warmup}_d + \mathit{maxDur}_d$: retire $d$ \\
  \> \> $r \gets d.\textsf{check}()$ \\
  \> \> $c_d \gets r.\mathit{triggered}\ ?\ c_d + 1 : 0$ \\
  \> \> \textbf{if} $c_d \geq \mathit{consecutive}_d$: cancel; \textbf{return}
  \end{tabbing}
  \caption{Monitor loop. Each detector has its own warm-up, poll interval,
  consecutive-violation threshold, and monitoring duration. Clear checks reset the
  violation count, so only sustained violations terminate an experiment.}
  \Description{Pseudocode for polling dependency detectors. Each detector is
  skipped during warm-up, checked on its own schedule, and triggers cancellation
  after a configured number of consecutive violations.}
  \label{fig:monitor}
  \end{figure}

Figure~\ref{fig:monitor} shows the loop. Detectors are skipped during warm-up,
  while deployment and replay are still settling, and retired after their
  monitoring window so long-running experiments do not keep unnecessary checks
  active. Requiring consecutive violations avoids terminating experiments on
  single transient excursions.

The platform uses two detector classes. \emph{Dependency-health detectors}
  observe whether a shared downstream dependency remains healthy enough for the
  experiment to continue. They read the dependency's own counters and compare them
  against a baseline captured at the start of the monitoring window, as well as
  against absolute limits. For example, one detector tracks unexpected worker
  restarts in the prediction service; another tracks downstream exception and
  invalid-result counts. These detectors prevent an experiment from continuing
  when the downstream environment is unstable, but they do not by themselves prove
  that the experiment caused the instability.

  \emph{Experiment-attributed detectors} provide that narrower signal when the
  downstream records carry enough caller identity. The malformed-request detector
  resolves the control and treatment jobs for the running experiment, then queries
  the prediction service's error records for requests originating from those jobs.
  A violation is therefore evidence about the monitored experiment rather than
  about the shared dependency as a whole. This allows the platform to terminate one
  faulty experiment without interrupting unrelated validation.

  Two limits follow from this design. First, polling introduces detection latency:
  a condition must occur after warm-up, be observed on a poll, and persist for the
  configured number of consecutive checks. Monitors therefore bound how long a
  condition continues; they do not prevent instantaneous failures, such as an
  invalid request that leaves an accelerator-backed worker unusable. Preventing
  that class belongs at the dependency boundary, through input validation before a
  request is accepted.

  Second, attribution is only as precise as the identity available in downstream
  records. If a downstream service records errors at batch granularity, a malformed
  request can implicate the jobs that contributed to the batch rather than only the
  single originating request. Section~\ref{sec:eval} accounts for this when
  reporting monitor terminations, and Section~\ref{sec:limitations} states the
  general limit: attribution is strongest at the immediate caller and weakens with
  depth.

\subsection{Bounded exposure: traffic pinning}
\label{sec:guardrails:pinning}

Prediction is difficult to isolate because the service hosts many model
deployments on accelerator-backed workers. Maintaining a representative isolated
tier for every model and variant would require reserved accelerator capacity.
Monitoring still applies, but it can react only after a fault becomes visible.

\emph{Traffic pinning} limits what an experiment can reach before detection.
Under normal routing, calls from an experiment are sent to one designated shard
instead of being distributed across every shard serving the model. A malformed
request that crashes a worker or corrupts device state is therefore confined to
the designated shard. Other shards remain available for serving traffic.

Pinning trades breadth of exposure for concentrated load. It is useful for
crash-class failures, where impact depends on how many workers receive a malformed
request. It is less suitable for resource contention because concentrating replay
traffic can make the designated shard a straggler. Dependency-health monitoring
therefore remains necessary alongside pinning.

\subsection{Applying the guardrails}
\label{sec:guardrails:composition}

The mechanisms are selected from properties of each downstream dependency. When
a representative copy of the dependency can be maintained with practical resource
demand, the workload routes replay to an isolated tier. When a dependency must
remain shared, the workload attaches health monitors wherever suitable downstream
signals are available. If the dependency is partitioned and its main failure mode
is local to a worker or shard, traffic pinning adds a bound on fan-out.

These choices address different failure classes. Isolation removes experimental
traffic from a shared request path. Monitoring bounds how long a visible violation
can continue. Pinning bounds the workers reached before detection. Lifecycle
effects and writes to shared state require separate controls, such as suppressing
the operation, tagging experimental traffic, or limiting corpus age. No one
guardrail covers every effect described in Section~\ref{sec:platform:boundary}.

Coverage is the combination of these choices rather than a property of one
mechanism. The guardrails also do not make experiments independent. Concurrent
experiments can share isolated tiers, pinned workers, and dependency-health
signals. Section~\ref{sec:limitations} returns to the resulting interference and
attribution limits.

\section{Experimentation Platform}
\label{sec:platform}

This section describes the platform's experiment topologies, execution model,
change inputs, lifecycle triggers, production connectivity, and replay behavior.

\subsection{Experiment topologies}
\label{sec:platform:topologies}

The platform supports two complementary topologies for collecting and evaluating
experiment evidence. They differ in how checks are defined and when they run;
neither changes which downstream dependencies the service under test (SUT) can
reach.

\begin{figure*}[t]
\centering
\begin{minipage}[t]{0.55\textwidth}
  \centering
  \includegraphics[width=\linewidth]{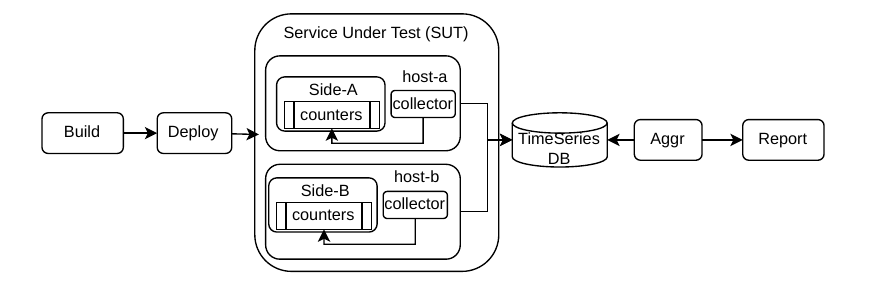}
  \smallskip
  {\small\textbf{(a)} Counter aggregation}
\end{minipage}
\hfill
\begin{minipage}[t]{0.41\textwidth}
  \centering
  \includegraphics[width=\linewidth]{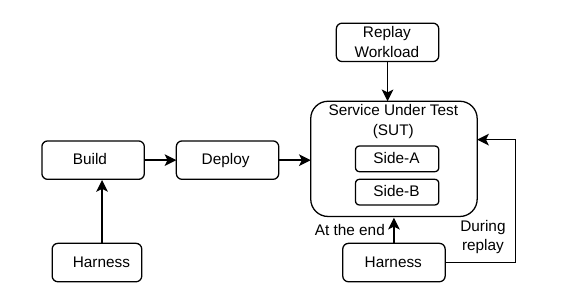}
  \smallskip
  {\small\textbf{(b)} Harness-driven checks}
\end{minipage}
\caption{Experiment topologies. (a) Collectors aggregate counters from paired SUT
instances and their hosts. (b) A programmable harness checks one SUT instance or
compares paired instances; the paired form is shown.}
\Description{Two experiment topologies are shown side by side. In the
counter-aggregation topology, collectors aggregate service and host counters from
paired instances. In the harness-driven topology, a programmable harness applies
checks during and after execution to either one SUT instance or paired instances.}
\label{fig:topologies}
\end{figure*}

Figure~\ref{fig:topologies}(a) shows the counter-aggregation topology. Control
and treatment versions of the SUT run as paired sides. A host-local collector
reads service-specific counters from endpoints exposed by each side and
resource-usage counters from each host. The samples are written to a time-series
store, aggregated across trials, and converted into a regression report. This
topology supports broad comparisons of resource use, latency, reliability, and
service-specific counters.

Figure~\ref{fig:topologies}(b) shows the harness-driven topology. A programmable
harness can evaluate either one SUT instance or paired control and treatment
instances. Single-instance tests include fuzzing and other checks that determine
whether an input violates a required property. Paired tests can compare
service-specific outputs directly, such as ranking order or normalized discounted
cumulative gain (NDCG). The harness can evaluate conditions during execution,
perform final checks when the test ends, and terminate the test after a sustained
violation.

The two topologies can also be combined. The dependency monitors in
Section~\ref{sec:guardrails:monitoring} read counters from shared downstream
dependencies and use the programmable harness to evaluate them while an experiment
runs. Counter collection supplies the health signals; the harness supplies the
polling, threshold, and termination logic. In either topology, calls leaving the
SUT may still reach shared production dependencies, so the guardrails in
Section~\ref{sec:guardrails} apply independently of how evidence is evaluated.

\subsection{Experiment execution}
\label{sec:platform:substrate}

Each workload runs as several separately allocated paired trials. Every trial has
a warm-up period followed by a measurement window. Control and treatment are
matched by logical host type, processor family, kernel version, operating system,
hardware fingerprint, and region. Trials that fail for infrastructure reasons may
be replaced up to a configured limit, while attempted, replaced, and excluded
trials remain recorded for diagnosis.

\subsection{Change delivery}
\label{sec:platform:delivery}

Two versioned stores matter. The source monorepo contains service code,
libraries, build definitions, and shared infrastructure. A separate repository
contains runtime configuration, including service tunables and traffic controls.
This creates two classes of change with different dependency signals.

A source change can alter executable behavior, dependencies, or resource use, and
its declared reach is visible in the build graph. A configuration change can alter
runtime behavior without changing a build target, so build-graph reachability does
not apply. Section~\ref{sec:triggering} describes how trigger selection handles
this difference.

Builds are revision-scoped. A package is produced from a target graph evaluated at
one source revision, with first-party dependencies drawn from that snapshot.
Source control is provided by Sapling~\cite{goode2022sapling}.

\subsection{Lifecycle triggers}
\label{sec:platform:triggers}

\begin{figure}[t]
  \centering
  \includegraphics[width=\columnwidth]{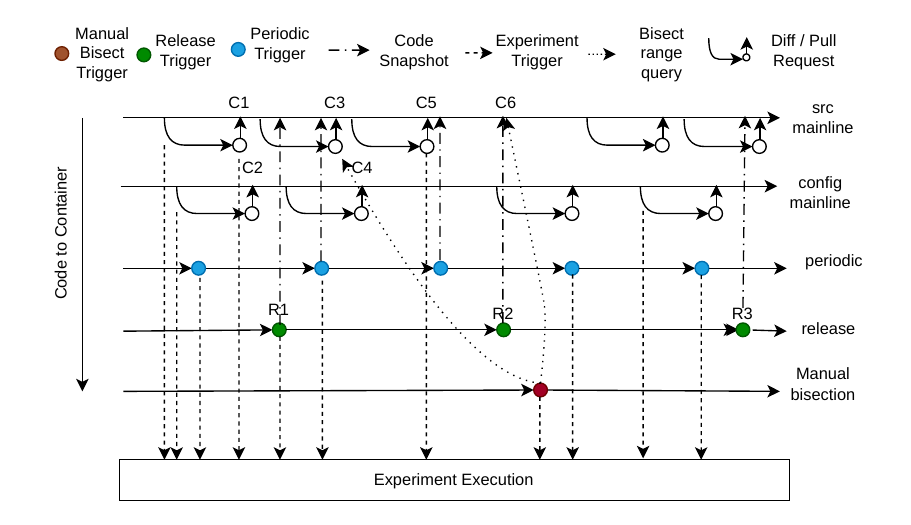}
  \caption{Experiment triggers across the change lifecycle, including bisection
  of a candidate revision range.}
  \Description{Source and configuration changes, periodic validation, release
  qualification, and manual bisection submit work to a common experiment
  executor.}
  \label{fig:triggers}
  \end{figure}

The platform is invoked by several lifecycle workflows. These workflows differ
  in which revision they test, how precisely a regression can be attributed, how
  much latency a developer or release process can tolerate, and how much shared
  downstream exposure they create. Figure~\ref{fig:triggers} shows how source,
  configuration, periodic, and release triggers feed the same experiment
  executor. When a later-stage run detects a regression, bisection can select
  intermediate revisions from the candidate range and submit additional runs.

  In Figure~\ref{fig:triggers}, curved arrows represent proposed source or
  configuration changes joining their mainlines. Dash-dotted paths create a
  revision-scoped code snapshot, and dashed paths submit that snapshot for
  experiment execution. Colored markers distinguish periodic, release, and
  manually initiated bisection runs; $C$ and $R$ labels denote successive changes
  and releases. A bisection range query identifies the
  revisions between a known-good and a known-bad result; the manual trigger then
  selects intermediate revisions from that range for additional experiments.
  The vertical direction follows the progression from code to an executable
  container.

  \begin{table}[t]
  \centering
  \small
  \setlength{\tabcolsep}{3pt}
  \begin{tabular}{@{}llllr@{}}
  \toprule
  Trigger & Blocking role & Attribution & Exec. share & Trial-time \\
  \midrule
  Diff-time & policy gate & one change  & 62\% & 62\% \\
  Land-time & detection   & range       & 10\% & 12\% \\
  Periodic  & calibration & n/a         & 9\%  & 13\% \\
  Release   & promotion   & release     & 18\% & 12\% \\
  Other     & tooling     & ad hoc      & 1\%  & $<$1\% \\
  \bottomrule
  \end{tabular}
  \caption{Lifecycle triggers for one ranking service over a representative
  measurement window.}
  \label{tab:triggers}
  \end{table}

  \emph{Diff-time} validation runs before submission, when a regression usually
  maps to one proposed change. \emph{Land-time} validation re-tests the mainline
  revision, which may contain a range of changes. \emph{Release} validation tests
  the artifact prepared for promotion, providing the strongest release fidelity
  but the weakest attribution.

  \emph{Periodic} A/A experiments run independently of a source or configuration
  change. They remain in Table~\ref{tab:triggers} to account for platform volume.

Execution share counts experiment runs, while trial-time share sums their trial
  duration. The similar diff-time shares indicate near-average duration. Land-time
  and periodic executions run longer on average, while release executions run for
  less time. Counts describe trigger volume; trial time approximates platform
  demand.

\subsection{What reaches production}
\label{sec:platform:boundary}

  Two boundaries have to be distinguished. The first is the boundary around the
  SUT itself. On the ingress side, the SUT serves replayed requests rather than
  user requests, and no product surface depends on the availability of the
  experimental deployment. A crash or local performance regression in the SUT
  therefore affects the experiment, not the user-facing request path.

  The second boundary is the SUT's egress. A downstream call leaving the SUT can be
  handled in three ways: redirected to an isolated tier reserved for
  experimental traffic; confined to a bounded subset of workers behind the
  dependency; or sent to the shared downstream dependency without redirection.
  Section~\ref{sec:guardrails} describes the first two mechanisms and when each
  applies. The third case is what makes production-connected experimentation
  different from fully isolated pre-production testing.

  When downstream calls remain connected, an experiment can affect shared
  dependencies in several ways.

  \begin{itemize}[leftmargin=1.2em,itemsep=1pt,topsep=2pt]
  \item \textbf{Capacity.} Replayed requests consume downstream capacity. Because
    ranking requests fan out, downstream demand can exceed the experiment's request
    rate and concurrent experiments compound it.

  \item \textbf{Crash or corruption.} A candidate change can emit input that a
    shared dependency mishandles, or replay can exercise a latent dependency
    defect. One malformed request can leave a worker unavailable, so impact need
    not be proportional to traffic share.

  \item \textbf{Lifecycle side effects.} Some downstream interactions affect
    resource accounting even when they do not write user-visible state. For
    example, replaying captured traffic for a stale model refreshed its usage
    reference count and delayed reaping. Such calls may not alter a user response,
    but they can change resource lifecycle decisions. Workloads must suppress,
    redirect, tag, or account for these interactions.

  \item \textbf{Observability.} Downstream services may count replayed requests in
    load, autoscaling, dashboard, or alerting signals unless caller identity
    distinguishes experiment traffic. Concurrent experiments can therefore also
    perturb a dependency-health baseline.
  \end{itemize}

  Experiment-local monitoring observes the control and treatment deployments. It
  does not, by itself, capture these downstream effects, which is why the platform
  needs the guardrails in Section~\ref{sec:guardrails}.

\subsection{Replay fidelity and its limits}
\label{sec:platform:fidelity}

Three properties of the replay path shape what the measurements mean: sampled
  recording, paired replay, and bounded recency.

  For each tier, the platform records traffic from a small number of serving
  tasks. Those tasks sample incoming requests uniformly. In the workloads we study,
  sampling rates range from $10^{-5}$ to $6.74\times 10^{-5}$, corresponding to
  approximately one in 100{,}000 to one in 14{,}800 eligible requests. The recorded
  corpus is therefore a sample of recent tier traffic, not a complete trace.

  During an experiment, both control and treatment receive the same recorded
  requests. For each request, the load generator sends the request to both sides,
  randomizing which side starts first. Any unusually heavy or light request
  therefore appears in both arms, reducing request-mix noise.

The corpus is drawn from a recent lookback window, typically one to four hours.
  This bounds how stale replayed accesses can be, reducing lifecycle side effects
  such as those described in Section~\ref{sec:platform:boundary}. Replay remains an
  approximation: it preserves sampled request contents, not full serving
  concurrency or causal user sessions.

  Replay rates are set to exercise the SUT at a high but controlled utilization.
  This leaves headroom against overload while making resource regressions visible
  in latency. The measured latency change is useful for deciding whether the
  candidate regresses the service, but it is not a claim that the same latency
  change would appear under every serving load.

  Pairing does not make the two sides independent. Both can call the same downstream
  dependency, so a treatment-side effect can also influence measurements observed
  by control. Dependency monitors limit harmful interference, but they do not remove
  this statistical coupling.

\section{Dependency-Aware Triggering}
\label{sec:triggering}

The platform validates two broad classes of changes: source-code changes that
  affect built artifacts, and configuration changes that affect runtime behavior
  without necessarily changing the binary. Every experiment consumes capacity and
  can add load to shared downstream dependencies. Trigger selection is therefore a
  containment mechanism, not only an efficiency optimization: a workload should run
  for a change only when the change can affect the artifact or runtime behavior
  that workload validates.

For source-code changes, the platform uses dependency reachability. A workload is
  eligible only when the changed files can reach the package target used by
  that workload through the declared build graph. This avoids running experiments
  for services or tiers outside the change's dependency closure.

Configuration changes require a different signal. Prefix-path filters provide a
  coarse first pass, but a prefix match can include stale, unused, or unrelated
  configuration. It also cannot tell which tiers actually consume a setting at
  runtime. The platform therefore maintains a runtime consumption index keyed by
  service and configuration name. A configuration change triggers a workload when
  that service has recently read the changed configuration at runtime. This ties
  experiment selection to observed use, rather than to repository layout alone.

Because the consumption index operates at configuration-file granularity, we also
  evaluate a setting-level refinement that uses recent per-setting reads and keeps
  the file-level decision when evidence is missing.

The platform also supports priority sampling across workloads. This mechanism is
  different from reachability filtering. Reachability attempts to exclude only
  workloads that the change cannot affect; sampling deliberately runs only a
  fraction of otherwise relevant diff-time workloads for lower-priority tiers, with
  later lifecycle stages providing additional coverage. This trades earlier
  coverage for lower resource demand and less exposure to shared downstream
  dependencies. In the measured rollout, source dependency filtering reduced
  diff-time workload executions by 25\% without changing the artifacts evaluated
  by the remaining experiments.

\section{Evaluation}
  \label{sec:eval}

  We first evaluate the guardrails in aggregate and then report measurements for
  each mechanism. We then evaluate dependency-aware triggering. Each result states
  its measurement scope and distinguishes observed outcomes from
  configuration-level guarantees.

  \subsection{Aggregate guardrail outcomes}
  \label{sec:eval:aggregate}

\begin{figure}[t]
  \centering
  \includegraphics[width=\columnwidth]{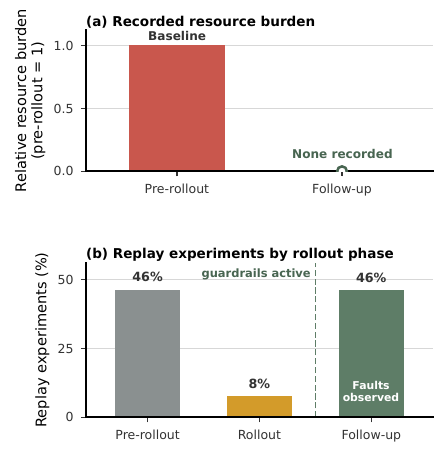}
  \caption{Aggregate outcomes across the incremental guardrail rollout. (a)
  Resource burden from targeted failure observations is normalized to the
  pre-rollout reference. Faults were still observed after rollout, but no burden
  was recorded during follow-up. (b) Pre-rollout and follow-up each contain 46\%
  of measured replay experiments, while the rollout period contains the 8\% run
  under partial coverage.}
  \Description{Two panels compare periods before, during, and after guardrail
  rollout. The first shows resource burden from targeted failure observations
  relative to a pre-rollout reference, with no burden recorded during follow-up
  despite observed faults. The second shows the share of replay experiments in
  each period, including the interval of partial guardrail coverage.}
  \label{fig:guardrail-burden}
  \end{figure}

  Figure~\ref{fig:guardrail-burden} aggregates the failure observations used in
  this study by position relative to the combined guardrail rollout. We use a
  derived resource-burden measure associated with each observation. The figure
  compares the combined burden from malformed-input and latent downstream-defect
  observations targeted by the guardrails. Absolute values and per-class shares
  are not reported, and records without a reported value are excluded.

  The guardrails were introduced incrementally rather than enabled as one atomic
  change. We therefore divide the measurement window into three periods.
  \emph{Pre-rollout} precedes the mechanisms evaluated here, \emph{rollout}
  contains experiments with partial guardrail coverage, and \emph{follow-up}
  begins after all three mechanisms were active on the covered paths. This avoids
  assigning experiments run under partial coverage to either endpoint.

  The replay volume is nearly balanced across the endpoint periods: 46\% of the
  measured experiments ran before rollout, 8\% during rollout, and 46\% during
  follow-up. Faults continued to be observed during the 81-day follow-up. The
  guardrails do not stop a candidate from producing faulty behavior or prevent a
  latent fault in a production dependency from being exercised. Instead,
  isolation redirected the effects, pinning bounded the workers they could reach,
  and monitors stopped experiments after violations became visible. The derived
  resource-burden measure recorded no burden during follow-up. This is an
  observed containment outcome, not a causal estimate that the guardrails reduced
  the underlying fault rate.

  \subsection{Per-guardrail effectiveness}
  \label{sec:eval:guardrails}

  \paragraph{Selective isolation.}
  Over a representative multi-day window, isolated ranking tiers handled billions
  of replayed requests per day. Without this redirection, those requests would
  have reached the corresponding shared ranking tiers. The isolated tiers require
  less than one percent of their provisioned worker capacity. Request volume is
  estimated from sampled routing records; the capacity comparison uses worker
  counts. The result applies to the covered ranking paths.

\begin{figure}[t]
  \centering
  \includegraphics[width=\columnwidth]{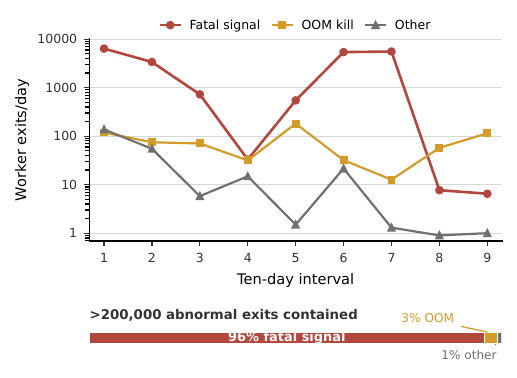}
  \caption{Abnormal worker exits in isolated ranking tiers over 90 days. The
  upper panel reports exits per day across nine ten-day intervals on a logarithmic
  scale; the lower bar summarizes the total as 96\% fatal-signal exits, 3\% OOM
  kills, and 1\% other exits. Planned updates and preemptions are excluded.}
  \Description{A line chart shows fatal-signal, out-of-memory, and other worker
  exits per day across nine ten-day intervals. A stacked bar shows that fatal
  signals account for 96 percent of more than 200,000 abnormal exits.}
  \label{fig:isolated-crashes}
  \end{figure}

  To measure the failures contained by this isolation, we counted abnormal
  worker exits in the isolated ranking tiers over 90 days. We excluded planned
  updates and infrastructure preemptions and deduplicated reports by task
  instance. As Figure~\ref{fig:isolated-crashes} shows, the tiers contained more
  than 200{,}000 abnormal worker exits: 96\% were fatal-signal exits, 3\% were OOM
  kills, and 1\% were other exits. Routing replay traffic to these tiers prevented
  these failures from directly affecting production-serving workers. Many exits
  were repeated instances of the same crash loops, so the count measures contained
  worker failures rather than distinct defects.

  \paragraph{Dependency monitoring.}
  Experiment-attributed detectors terminated experiments that emitted malformed
  requests to a shared prediction service by matching downstream errors to the
  originating experiment.

  Over a separate three-month observation window, dependency monitors cancelled
  approximately 5\% of the service's experiments and accounted for about 87\% of
  all experiment cancellations. Of these monitor-initiated cancellations,
  approximately 81\% came from the dependency-health detector for
  prediction-worker crashes and 19\% from the experiment-attributed
  malformed-request detector.

  For response time, Table~\ref{tab:latency} uses a separate 16-day window and
  reports the interval between a detector first observing a violation and the
  experiment being terminated. For both detector classes, the median response is
  approximately one configured polling interval after the first observation.

  \begin{table}[t]
  \centering
  \small
  \setlength{\tabcolsep}{5pt}
  \begin{tabular}{@{}lrrrr@{}}
  \toprule
  Detector & $n$ & Median & p90 & Maximum \\
  \midrule
  Dependency health     & 534 &  62\,s &  64\,s &  69\,s \\
  Experiment-attributed &  14 & 121\,s & 121\,s & 364\,s \\
  \bottomrule
  \end{tabular}
  \caption{Interval from a detector's first observed violation to experiment
  termination.}
  \label{tab:latency}
  \end{table}

  The interval begins when the detector first observes the violation, not at fault
  onset. Dependency-health terminations are also not independent: one disturbance
  can terminate several experiments, so their count measures exposed experiments
  rather than distinct faults.

  \paragraph{Traffic pinning.}
  Pinning directs experiment requests to one designated shard rather than every
  shard serving the model. This bounds task-local failures to that shard, but not
  effects propagated through shared state. Because telemetry cannot establish how
  many workers would otherwise have failed, we report the routing bound rather
  than avoided impact.

\begin{figure}[t]
  \centering
  \includegraphics[width=\columnwidth]{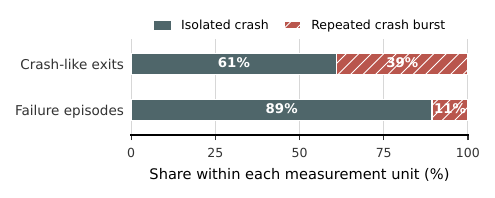}
  \caption{Crash concentration on pinned prediction tasks over 60 days. Crashes
  on the same task separated by at most ten minutes form one episode. Repeated
  bursts account for 39\% of crash-like exits but 11\% of failure episodes.}
  \Description{Two stacked bars compare isolated crashes with repeated crash
  bursts. Repeated bursts represent 39 percent of crash-like exits and 11 percent
  of failure episodes after crashes are grouped by task and time.}
  \label{fig:pinned-crash-bursts}
  \end{figure}

  We also measured how crashes recur on pinned prediction tasks. We identified
  the designated tasks from a representative one-day routing snapshot and examined
  their task-exit telemetry over the preceding 60 days. We excluded planned
  updates and infrastructure preemptions, deduplicated reports by task instance,
  and treated segmentation faults and OOM kills as crash-like exits. Consecutive
  crashes on the same task, separated by at most ten minutes, form one failure
  episode. As Figure~\ref{fig:pinned-crash-bursts} shows, 39\% of crash-like exits
  occurred in repeated bursts. After each burst is counted once, repeated bursts
  account for 11\% of failure episodes. This grouping prevents a crash loop from
  being counted as a sequence of independent containment events. The measurement
  characterizes recurrence on pinned targets; it does not estimate how many
  production workers would have failed without pinning.

  \subsection{Dependency-aware triggering}
  \label{sec:eval:triggering}

  Source-code selection uses build-graph reachability; configuration selection
  uses observed runtime consumption.

  In the measured rollout, source dependency filtering reduced diff-time workload
  executions by 25\%. The artifact, replay corpus, and trial configuration of the
  remaining experiments were unchanged. The reduction comes from avoiding
  workloads outside the expected dependency closure, not from shorter runs.

  An offline replay of 775 changed settings found that this refinement would
  suppress 72.3\% of the changes selected by file-level consumption. In a held-out
  final week, 1.7\% of suppressed settings were later observed as read by the SUT.
  These results evaluate the selection rule offline; the refinement is not enabled
  in the measurements reported here.

  The first result counts workload executions; the second counts changed settings,
  not executions or compute saved. Section~\ref{sec:limitations} discusses the
  coverage limits of both selectors.

\section{Related Work}
  \label{sec:related}

  \paragraph{Canary deployment.}
  Canary and progressive delivery systems expose a change to production traffic
  only after earlier validation. Canary Analysis Service, Kayenta, Gandalf, and
  Conveyor compare candidate and baseline behavior during limited rollout and
  control promotion~\cite{davidovic2018canary,kaul2018kayenta,li2020gandalf,
  grubic2023conveyor}. Open systems package the same pattern as automated
  progressive delivery~\cite{argo-rollouts,flagger,keptn-quality-gates}, while
  learned risk models increasingly inform promotion decisions
  \cite{abreu2025moving}. These systems limit exposure during rollout. Our setting
  instead reaches shared dependencies while a candidate is still under review,
  before rollout can provide feedback.

  \paragraph{Record-and-replay validation.}
  Record and replay improves input fidelity but does not determine how dependencies
  are handled. Liu et al. identify mocking points so replayed microservice requests
  execute against isolated dependencies~\cite{liu2022recordreplay}; AREX, Keploy,
  and GoReplay make different choices about capture scope and dependency
  stubbing~\cite{arex,keploy,goreplay}. Diffy sends the same requests to candidate
  and known-good instances and compares responses~\cite{diffy}, but response
  comparison does not itself isolate their downstream calls. Preproduction
  environments provide controlled connectivity from candidate deployments to live
  services~\cite{carroll2021preproduction}.

  ServiceLab is a published third-party system that provides paired replay,
  matched execution environments, counter comparison, and baseline-based
  regression detection~\cite{chow2024servicelab}. It also permits selected calls
  from the SUT to reach dependencies that handle production traffic, while leaving
  side-effect prevention to service owners. Our work addresses the complementary
  platform problem: systematic containment, monitoring, and attribution when
  experiments are not fully isolated.

  \paragraph{Production-scale testing and benchmarks.}
  Kraken drives live traffic to find resource bottlenecks, WSMeter evaluates
  warehouse-scale hardware with production workloads, and Treadmill provides
  precise open-loop load generation and tail-latency attribution
  \cite{veeraraghavan2016kraken,lee2018wsmeter,zhang2016treadmill}. DeathStarBench
  and TeaStore provide reproducible microservice topologies
  \cite{gan2019deathstarbench,vonKistowski2018teastore}. These systems evaluate an
  existing deployment or a controlled benchmark. Our concern is a candidate change
  whose replayed calls converge with production traffic at shared dependencies.

  \paragraph{Regression-test selection.}
  Test selection uses observed dependencies, commit pruning, or learned
  prediction~\cite{gligoric2015ekstazi,memon2017taming,
  machalica2019predictive}. Our source-change selector uses build-graph
  reachability, which is sound only relative to the declared dependency closure;
  runtime loading, generated code, and undeclared inputs require conservative
  fallback. Configuration state is absent from that graph, so we instead use
  observed runtime consumption. In our setting, selection reduces both experiment
  capacity and calls to shared dependencies, making it a reliability mechanism as
  well as a scheduling optimization.

  \paragraph{Containment and chaos engineering.}
  Chaos engineering perturbs production within a controlled blast radius and
  aborts when guardrail signals degrade~\cite{basiri2016chaos}. ChAP is the closest
  published analogue to our guardrail layer: it routes a fraction of production
  traffic to instrumented control and experiment clusters and terminates a run on
  metric deviation~\cite{basiri2019automating}. A chaos experiment specifies its
  perturbation, even when the resulting behavior is uncertain. Candidate code has
  a broader failure space, so our guardrails instead combine downstream health,
  experiment-attributed errors, selective isolation, and bounded fan-out.

  \paragraph{Failure attribution and tracing.}
  Distributed tracing systems propagate causal context across service boundaries,
  and dynamic tracing systems evaluate cross-service queries
  \cite{sigelman2010dapper,kaldor2017canopy,mace2015pivottracing}. Our monitors
  separate health detection from attribution. A health detector can observe a
  dependency at depth, while error attribution requires caller identity to survive
  the request path and is strongest at the immediate dependency. This distinction
  determines which failures can be assigned automatically to one experiment.

  \paragraph{Experiment analysis.}
  Online controlled experiments address related statistical questions at a
  different operating point: randomized users observed over long periods, rather
  than a small number of paired host trials
  \cite{kohavi2013online,kohavi2020trustworthy}. Our replay uses common random
  numbers~\cite{glasserman1992crn}: both sides receive the same corpus, removing
  request-mix differences by construction rather than estimating them afterward,
  as in covariate-based variance reduction~\cite{deng2013cuped}. Evaluating many
  counters raises the multiple-comparison problem~\cite{benjamini1995fdr}, while
  sensitivity to tail latency motivates precise paired measurements
  \cite{dean2013tail}.

  \paragraph{Matched placement.}
  Allocating control and treatment together is related to gang scheduling, where
  several tasks are useful only when placed as a unit
  \cite{ousterhout1982scheduling,feitelson1992gang}. Cluster managers expose
  placement constraints and multi-job allocation mechanisms
  \cite{verma2015borg,tang2020twine}. The matched environments in our platform use
  this scheduling capability as substrate; scheduling itself is not a contribution
  of this paper.

\section{Limitations}
  \label{sec:limitations}

  \paragraph{Attribution and containment boundaries.}
  Health monitors can observe configured dependencies, but attribution requires
  preserved caller identity or propagated experiment context
  \cite{kaldor2017canopy}. Traffic pinning confines task-local failures, not
  effects propagated through shared state. Isolated tiers provide a stronger
  boundary where faithful replication is practical; monitors act only after a
  violation becomes visible.

  \paragraph{Concurrent experiments.}
  Experiments sharing an isolated tier, pinned shard, or health signal remain
  coupled through contention and common disturbances. Detectors distinguish them
  only when downstream records preserve caller identity.

  \paragraph{Replay and selection coverage.}
  Replay preserves sampled requests, not the full distribution of concurrency or
  causal sessions, and selected recording tasks can omit traffic patterns
  (Section~\ref{sec:platform:fidelity}). Build-graph filtering covers declared
  dependencies; runtime filtering can miss settings not read during observation.
  Priority sampling also defers eligible workloads. Section~\ref{sec:eval:triggering}
  therefore measures executions avoided, not regressions missed.

  \paragraph{Generality.}
  Measurements focus primarily on one family of recommendation services. The design assumes
  placement constraints, service introspection, and downstream caller identity;
  other systems require equivalent scheduling, monitoring, and attribution.

\section{Conclusion}
\label{sec:conclusion}

Candidate changes receive realistic traffic and dependency behavior before they
enter the mainline, but production-connected execution also extends their effects
beyond the SUT. Some dependencies can be isolated, some can only be monitored,
and some require a bound on the workers that replayed traffic can reach.

The experimentation platform studied here combines isolated tiers, dependency monitoring,
and traffic pinning for these cases. Dependency-aware triggering supports the same
operating model by reducing experiment volume and downstream exposure. The broader
lesson is that early validation against shared serving dependencies requires
downstream effects to remain observable, attributable, and contained.

\bibliographystyle{ACM-Reference-Format}
\bibliography{servicelab_refs_merged}

\end{document}